\documentclass[aps,prl,superscriptaddress,twocolumn,showpacs,amsmath,amssymb,reprint,longbibliography]{revtex4-2}

\usepackage{graphicx}
\usepackage[percent]{overpic}
\usepackage{placeins}
\usepackage{dcolumn}
\usepackage{bm}
\usepackage[T1]{fontenc}
\usepackage[utf8]{inputenc}
\usepackage{xcolor}
\usepackage[colorlinks=true,allcolors=blue]{hyperref}

\usepackage{amsfonts}
\usepackage{mathptmx}
\begin{document}

\title{\texorpdfstring{N\'eel-Vector-Orientation Induced Direction-Robust Spin Filtering in Two-Dimensional Altermagnets}{N\'eel-Vector-Controlled Direction-Robust Spin Filtering in Two-Dimensional Altermagnets}}

\author{Xin Chen}
\affiliation{Thermal Science Research Center, Shandong Institute of Advanced Technology, Jinan 250100, Shandong Province, People's Republic of China}

\author{Jin Zou}
\affiliation{Faculty of Applied Sciences, Macao Polytechnic University, Macao SAR, 999078, People's Republic of China}

\author{Lipeng Song}
\affiliation{Thermal Science Research Center, Shandong Institute of Advanced Technology, Jinan 250100, Shandong Province, People's Republic of China}

\author{Wei Sun}
\affiliation{Thermal Science Research Center, Shandong Institute of Advanced Technology, Jinan 250100, Shandong Province, People's Republic of China}

\author{Yiwen Wu}
\affiliation{Thermal Science Research Center, Shandong Institute of Advanced Technology, Jinan 250100, Shandong Province, People's Republic of China}

\author{Luyao Zhu}
\affiliation{Thermal Science Research Center, Shandong Institute of Advanced Technology, Jinan 250100, Shandong Province, People's Republic of China}

\author{Xu Cheng}
\email{xchengab@sdu.edu.cn}
\affiliation{Institute for Advanced Technology, Shandong University, Jinan 250061, People's Republic of China}

\author{Duo Wang}
\email{duo.wang@mpu.edu.mo}
\affiliation{Faculty of Applied Sciences, Macao Polytechnic University, Macao SAR, 999078, People's Republic of China}

\author{Biplab Sanyal}
\email{biplab.sanyal@physics.uu.se}
\affiliation{Department of Physics and Astronomy, Uppsala University, Box 516, 751\,20 Uppsala, Sweden}

\date{\today}

\begin{abstract}
Whether an antiferromagnet can host direction-robust spin-polarized transport without a conventional spin-selective band gap remains a central challenge in antiferromagnetic spintronics. Here we establish a gapless, direction-robust spin-filtering mechanism in a compensated two-dimensional altermagnetic Weyl semimetal that requires neither a spin-selective band gap nor a large velocity contrast between spin projections. Using Janus monolayer Ta$_2$TeSeO as a realistic platform, we combine symmetry analysis with first-principles calculations, full-Brillouin-zone Wannier interpolation, and semiclassical transport. Rotating the N\'eel vector removes a unitary-mirror constraint and shifts one Weyl-cone pair away from its parent high-symmetry line. For an in-plane N\'eel vector, the residual $C_{2z}\mathcal T$ symmetry forbids the independent $\sigma_y$ mass that would open a local gap, allowing the reconstructed cones to shift in momentum while remaining gapless. Breaking unitary $C_{2z}$ simultaneously lifts the energy equivalence of the remaining mirror-pinned Weyl cones. The resulting coexistence of a metallic spin-projected manifold and a low-DOS Weyl-derived manifold produces a predominantly DOS-driven conductance imbalance. At charge neutrality and 20~K, the longitudinal conductivity polarization for $\mathbf n\parallel x$ remains positive for every in-plane current direction and ranges from $76.4\%$ to $82.0\%$. The degenerate in-plane magnetic anisotropy facilitates reversible switching between symmetry-related spin-filtering states using strain or weak anisotropic fields. This N\'eel-vector-driven symmetry mechanism provides a general route to direction-robust gapless spin filtering in compensated altermagnets.
\end{abstract}

\maketitle

In condensed-matter physics, simple control knobs such as electric fields or layer stacking angles are widely used to switch materials between qualitatively distinct electronic states. In magnetic systems, the direction of the order parameter is one of the most natural knobs \cite{PhysRevX.14.031037,li2025spinorbitcouplingdrivenchiralityswitching,duan2025neelvectorrashbasoc,zhiheng2025spinaxisdynamiclocking,refId0}, routinely used to control anisotropic magnetoresistance or spin-wave spectra. Much less explored is whether reorienting the magnetic order alone, without changing the lattice or composition, can drive a compensated antiferromagnet toward half-metal-like transport functionality.

Half-metals, in which one spin channel is metallic while the other is insulating, are central building blocks of spintronic devices because they provide 100\% spin-polarized carriers at the Fermi level. Established half-metals, however, are ferro- or ferrimagnetic and therefore generally carry a net magnetization. Such an AFM half-metal would offer a fully spin-polarized Fermi surface and zero net moment, combining the absence of stray fields and ultrafast antiferromagnetic dynamics. The concept of an AFM half-metal was introduced by van Leuken and de Groot in 1995 \cite{PhysRevLett.74.1171}, but the proposed mechanism hinges on a delicate cancellation of unequal antiparallel local moments. Transport analogues of this extreme limit have been proposed in compensated magnets, particularly in altermagnets \cite{PhysRevX.12.040501,PhysRevX.12.031042,PhysRevB.102.014422,smejkal2020crystal,PhysRevLett.132.236701,PhysRevX.12.011028,PhysRevLett.128.197202,Jungwirth2025,Zhu2024nature,ChenX2025}. These scenarios can yield highly spin-polarized transport without net magnetization.

Within a common-relaxation-time picture, the longitudinal spin polarization can be decomposed as
\begin{equation}
P_{\alpha\alpha}
=
\frac{G_{\alpha\alpha}^{+}-G_{\alpha\alpha}^{-}}
     {G_{\alpha\alpha}^{+}+G_{\alpha\alpha}^{-}},
\qquad
G_{\alpha\alpha}^{\pm}
\propto
\widetilde D_{\pm}\langle v_{\alpha}^{2}\rangle_{\pm}.
\label{eq:transport-decomposition}
\end{equation}
Here, \(\widetilde D_{\pm}\) is the projected DOS weight within the transport window and \(\langle v_{\alpha}^{2}\rangle_{\pm}\) is the corresponding velocity-squared moment. Large \(\lvert P_{\alpha\alpha}\rvert\) can arise from spectral exclusion, which suppresses \(\widetilde D\) for one projection through a gap or band-edge shift \cite{doi:10.1073/pnas.1715465115,PhysRevLett.134.116703,10.1063/5.0147450,PhysRevLett.133.216701,10.1093/nsr/nww026,doi:10.1002/adma.201102555,PhysRevB.110.L220402,10.1063/5.0250823,PhysRevB.107.214419,10.1063/5.0242426,PhysRevB.111.035437,Manc2021,Xu2025Alterpiezoresponse,v38b-5by1,PhysRevB.111.134429,10.1063/5.0252374,Khan2025}, or from strongly unequal velocity moments \cite{bf1n-sxdl,zhiheng2025spinaxisdynamiclocking}. The present mechanism is instead dominated by a phase-space contrast: one projection contains conventional metallic pockets with a relatively large \(\widetilde D\), whereas the other contains small Weyl-derived pockets with a much smaller \(\widetilde D\). Because their velocity moments remain comparable, the conductance imbalance is predominantly DOS driven. N\'eel-vector reorientation controls this contrast by changing the magnetic symmetry and redistributing the low-energy states between the projections. The resulting state is therefore a gapless spin filter rather than a strict half-metal.

\begin{figure}[hbtp]
\includegraphics[scale=1]{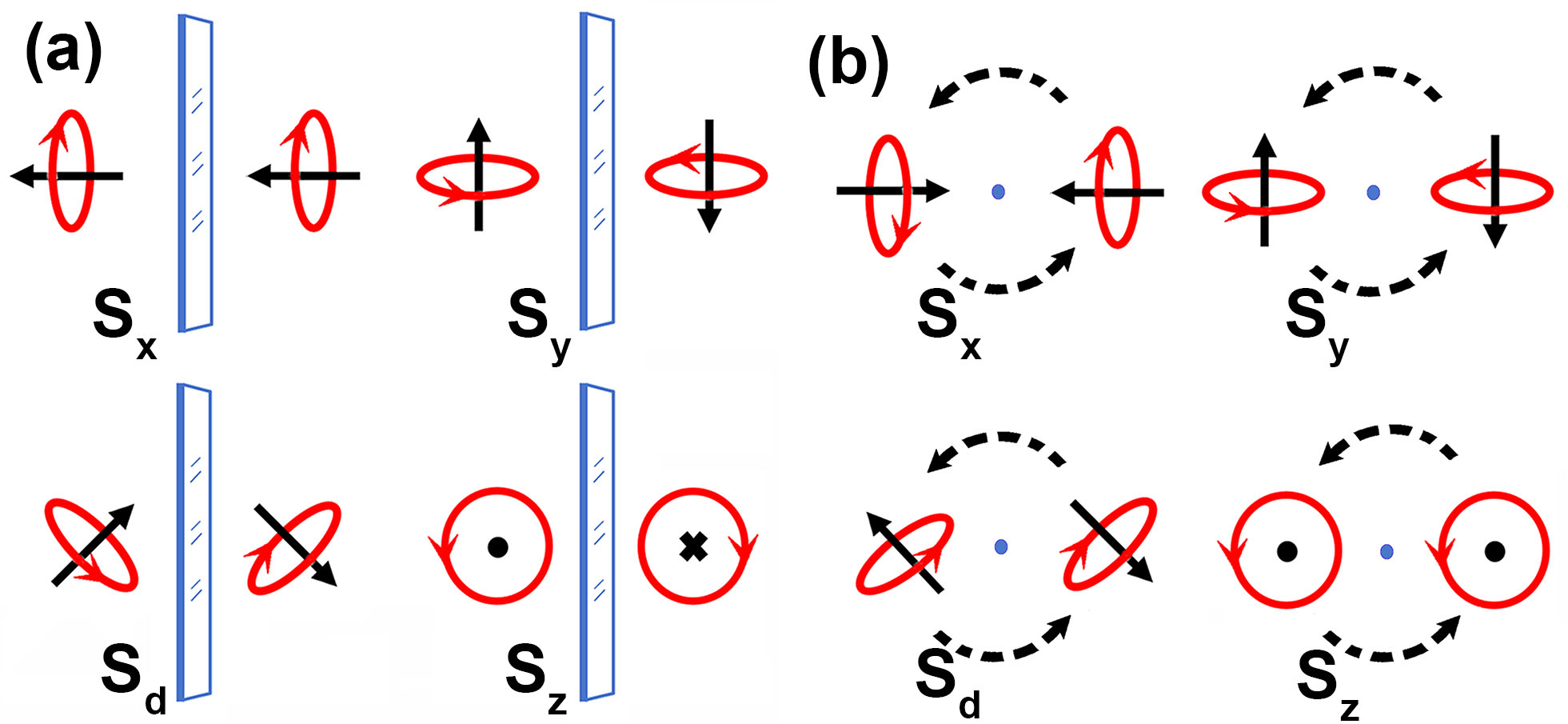}
\caption{(a) Mirror-symmetry operation $M_x$ and (b) twofold-rotation operation $C_{2z}$ applied to magnetic moments for N\'eel-vector orientations $\mathbf n\parallel x,y,z$, and $d\equiv(110)$. Black arrows indicate the moment directions, while red circular arrows depict the equivalent loop current, and blue lines mark the symmetry geometry.
}
\label{figure1}
\end{figure}

In this work, we propose a mechanism for direction-robust gapless spin filtering in two-dimensional altermagnetic Weyl semimetals and validate it using Janus monolayer Ta$_2$TeSeO as a realistic material platform. We show that the N\'eel-vector orientation acts as a symmetry switch that selectively reduces the magnetic space group. For an in-plane N\'eel vector, the symmetry reduction displaces one Weyl-cone pair from its parent high-symmetry line and lifts the energetic equivalence of the remaining cones. This reconstruction produces metallic pockets in one spin projection and a low-DOS Weyl-derived manifold in the other. Their comparable velocity moments identify the resulting conductance imbalance as predominantly DOS driven. At charge neutrality and 20~K, the calculated conductivity polarization ranges from \(76.4\%\) to \(82.0\%\) throughout the in-plane angular range. The symmetry-related \(x\)- and \(y\)-oriented N\'eel vectors interchange the dominant spin projection, enabling reversible switching between the two spin-filtering states.

To clarify how the N\'eel-vector $\mathbf n$ orientation affects the surviving magnetic symmetries, we briefly recall the transformation rules of an axial vector under mirror and twofold rotation operations (Fig.~\ref{figure1}). As an example, let the mirror plane be $yz$ ($M_x$), the axial-vector transformation is
\begin{equation}
M_x:(S_x,S_y,S_z)\mapsto(S_x,-S_y,-S_z),
\label{eq:axial_mirror}
\end{equation}
Rotations act on axial and polar vectors identically, so
\begin{equation}
C_{2z}:(S_x,S_y,S_z)\mapsto(-S_x,-S_y,S_z).
\label{eq:axial_rotation}
\end{equation}
Thus, $\mathbf{n} \| \hat{x}$ preserves the unitary $M_x$ but breaks $M_y$, which survives as the magnetic mirror $M_y\mathcal T$. For $\mathbf n\parallel y$ the assignments are interchanged. For an in-plane N\'eel vector, $C_{2z}$ reverses the moment and must be combined with time reversal to remain a symmetry. Thus, the composite antiunitary $C_{2z}\mathcal T$ is preserved. For $\mathbf n\parallel z$, $C_{2z}$ is retained, whereas for $\mathbf n$ along the diagonal $d\equiv(110)$, neither in-plane mirror remains unitary and unitary $C_{2z}$ is also lost in a fixed magnetic domain. {These orientation-dependent symmetries determine which parent crossings remain pinned to high-symmetry lines and which node energies remain symmetry related.}

\begin{figure}[hbtp]
\centering
\includegraphics[scale=1.0]{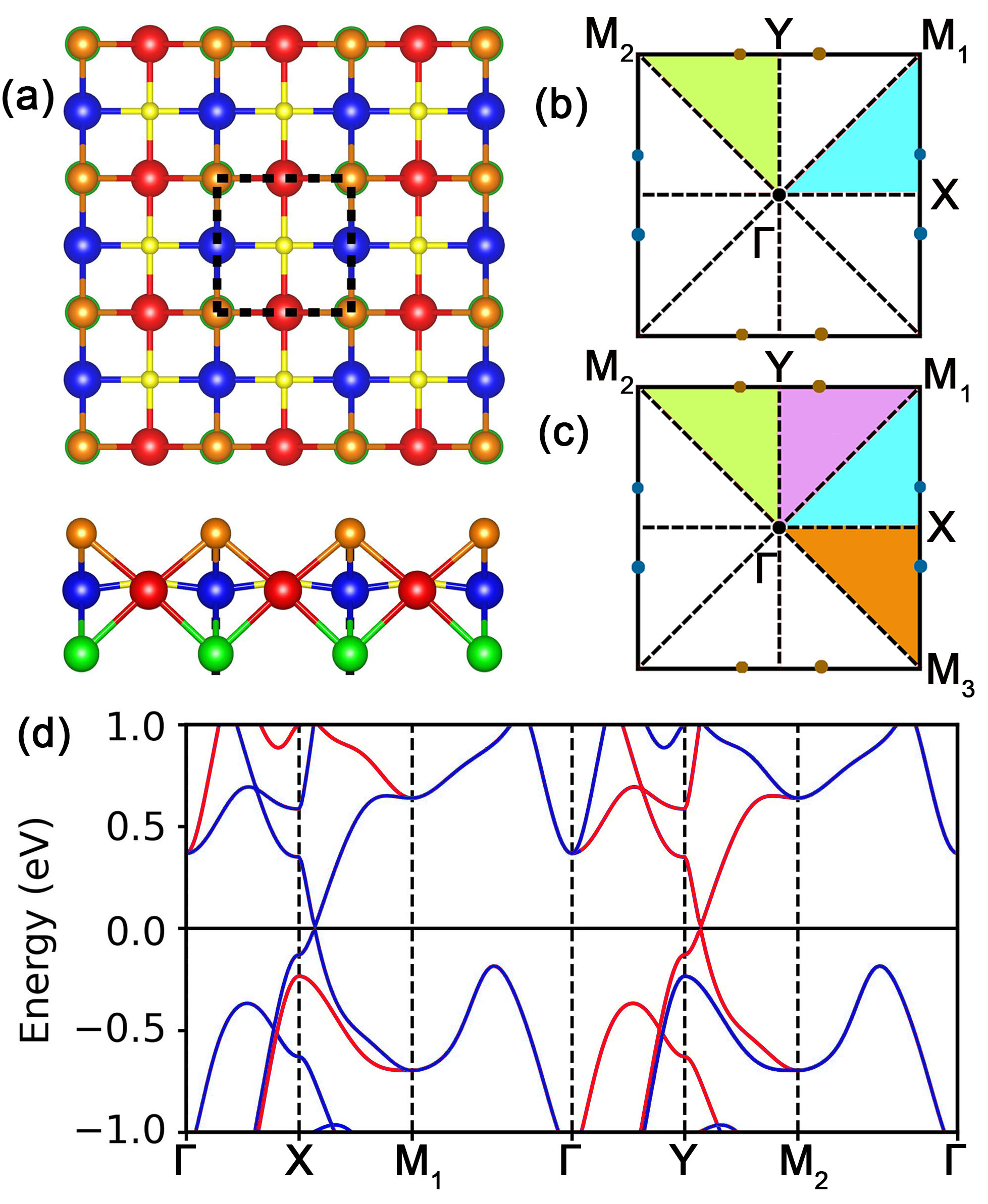}
\caption{(a) Top and side views of monolayer Ta$_2$TeSeO. Yellow, red, blue, orange, and green spheres denote O atoms, spin-up-polarized Ta atoms, spin-down-polarized Ta atoms, Se atoms, and Te atoms, respectively. (b,c) The first BZ of Ta$_2$TeSeO is shown with its irreducible sectors without and with SOC, respectively. (d) The band structure without SOC is shown. The up- and down-spin blocks are depicted in red and blue, respectively.}
\label{figure2}
\end{figure}

\begin{figure*}[hbtp]
\centering
\includegraphics[scale=0.9]{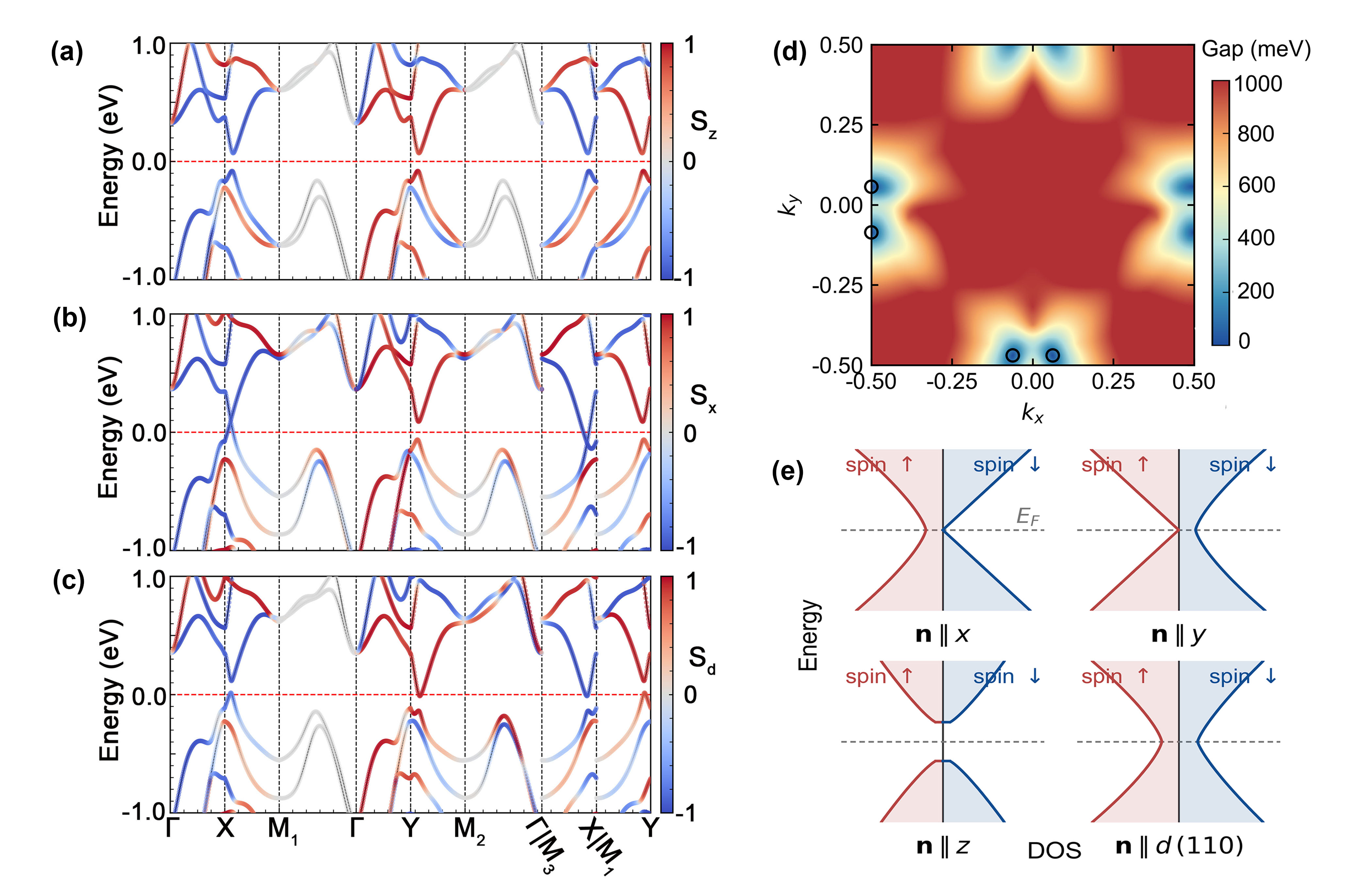} 
\caption{N\'eel-vector-dependent electronic reconstruction in Ta$_2$TeSeO. Spin-projected electronic band structures of Ta$_2$TeSeO for the magnetic axis aligned along (a) the out-of-plane $z$ direction, (b) the in-plane $x$ direction, and (c) the in-plane diagonal $(110)$ direction; the spin projections are depicted in red and blue, respectively. (d) Wannier-interpolated direct gap $E_{17}(\mathbf k)-E_{16}(\mathbf k)$ over the centered first BZ at $k_z=0$. Open circles mark refined low-gap structures. (e) Qualitative summary of the spin-projected spectral-weight redistribution; the corresponding DFT results are provided in the SM~\cite{SM}.}
\label{figure3}
\end{figure*}

One mirror remains unitary and protects the crossing on its mirror-invariant line. Once the corresponding unitary mirror is lost, both the local-mass and momentum-shift terms become symmetry-allowed, while the remaining magnetic symmetries further constrain the local Hamiltonian. The resulting reconstruction is therefore resolved over the full two-dimensional Brillouin zone. The complete axial-vector derivation and the orientation-dependent magnetic-symmetry classification are provided in the Supplemental Material (SM)~\cite{SM}.

Our calculated structure and compensated antiferromagnetic configuration of Janus Ta$_2$TeSeO are shown in Fig.~\ref{figure2}(a). In the nonmagnetic limit the monolayer adopts the symmorphic space group $P4mm$ (No.~99), whose point group is $C_{4v}$. The horizontal mirror $M_z$ is absent due to the Janus polarity. First-principles total-energy calculations identify the compensated antiferromagnetic configuration as the ground state. Calculated inter-site magnetic exchange parameters $J_{ij}$, evaluated using a dense $500\times500\times1$ $k$-point mesh, exhibit long-range sign oscillations with pair distance, most prominently along the $[110]$ Ta--Ta bonding direction. This directional behavior is consistent with an anisotropic RKKY-like contribution mediated by the itinerant states near the Fermi level \cite{PhysRevB.64.174402,PhysRevLett.116.217202}. Computational details, exchange-interaction calculations, Monte Carlo simulations, and structural and dynamical stability tests are given in the SM~\cite{SM}.

Explicit first-principles calculations show that the in-plane N\'eel-vector rotation barrier is on the order of $10~\mu\mathrm{eV}$ per unit cell (see the SM). The fully self-consistent noncollinear SOC state carries a small residual magnetic moment of $0.0158\,\mu_{\rm B}$ per unit cell and is therefore nearly, rather than exactly, compensated.

A small SOC-induced moment is compatible with antiferromagnetic and altermagnetic order and can arise from weak spin canting or unequal spin and orbital responses on opposing sublattices \cite{PhysRevB.71.184434,Milivojevic2024RuF4,PhysRevLett.134.196703}. In Ta$_2$TeSeO, the surviving magnetic symmetries allow a uniform moment only parallel to the in-plane N\'eel vector: $M_x$ for $\mathbf n\parallel x$ and $M_y$ for $\mathbf n\parallel y$. SOC also mixes spin, and the full-BZ spectrum contains no globally insulating spin projection because the Weyl crossings remain. Exact compensation is therefore not enforced by separate integer occupations of two spin channels. The residual moment and the transport polarization have distinct origins: the former reflects weak relativistic uncompensation, whereas the latter is governed primarily by the projected-DOS contrast.

Because the residual moment follows the N\'eel orientation, it provides a linear Zeeman coupling that allows an in-plane magnetic field to lift the degeneracy between the $x$- and $y$-oriented states and select the corresponding spin-filtering state. A related coupling between a weak moment and altermagnetic order has recently enabled field-assisted domain selection in MnTe \cite{Liu2026MnTeSwitching}. The small in-plane anisotropy makes such field selection energetically plausible, although a quantitative switching threshold also depends on domain formation and finite-temperature dynamics. Weak in-plane strain may likewise select between the two spin-filtering states. The coupling between the lattice and the N\'eel vector further suggests that circularly polarized light could provide an ultrafast, contact-free control knob \cite{Ueda2023,PhysRevLett.134.196906,PhysRevB.110.094401}. More generally, a small relativistic uncompensation need not undermine altermagnetic spin filtering: it can add a direct field-control handle while the transport response remains governed by the orientation-dependent band structure.

In the SOC-free limit, Ta$_2$TeSeO factorizes into two spin blocks. Each block realizes a pair of mirror-pinned linear crossings on the Brillouin-zone boundary [Figure~\ref{figure2}(b,d)]. Within each spin block, a minimal $k\cdot p$ Hamiltonian near the $i$th parent crossing reads
\begin{equation}
\begin{aligned}
H_{0,i}(\mathbf q)=&\,\varepsilon_{W,i}^{(0)}\sigma_0
+v_{i,\parallel}q_{\parallel}\sigma_x
+v_{i,\perp}q_{\perp}\sigma_z,\\
&M_i\sigma_xM_i^{-1}=\sigma_x,
\qquad M_i\sigma_{y,z}M_i^{-1}=-\sigma_{y,z},
\end{aligned}
\label{eq:parent_kp}
\end{equation}
where $\mathbf q$ is measured from the parent node and $q_{\parallel}$ ($q_{\perp}$) is parallel (perpendicular) to the mirror-invariant line. The Pauli matrices act in the two-band orbital subspace, not in real-spin space. The only momentum-independent mass term $m_i\sigma_y$ that can hybridize the crossing states anticommutes with both kinetic terms and opens a local gap \cite{Hsieh2012}. It is odd under the corresponding unitary mirror and is therefore forbidden by symmetry \cite{PhysRevB.88.241303}.

\begin{figure*}[hbtp]
\centering
\includegraphics[scale=1]{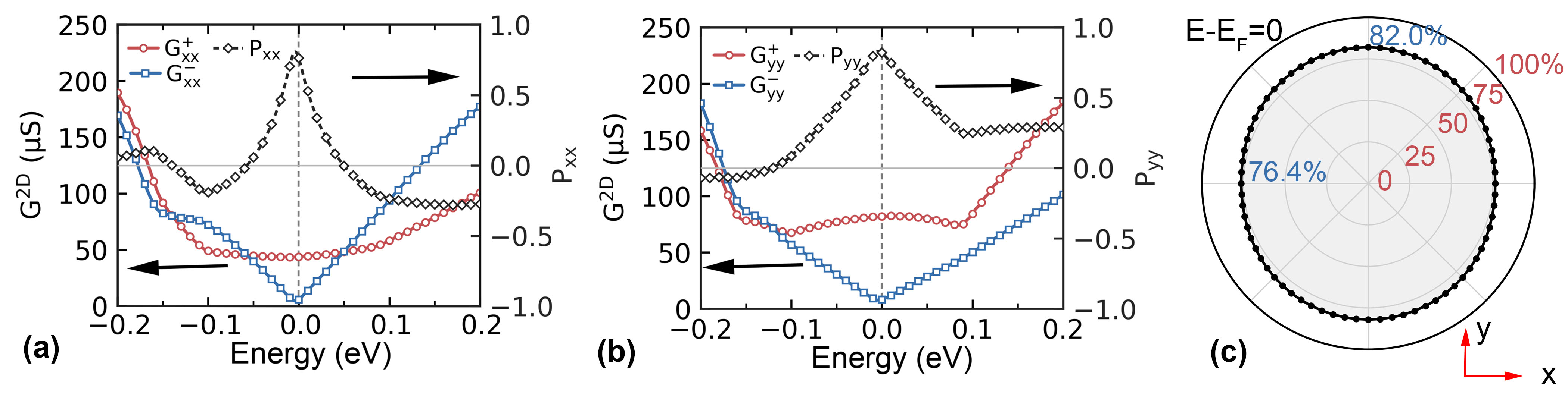}
\caption{Spin-projected longitudinal transport for $\mathbf n\parallel x$ at $T=20$~K using $\tau=10$~fs. Panels (a) and (b) show the $xx$ and $yy$ responses. Red open circles and blue open squares denote $G_{\alpha\alpha}^{+}$ and $G_{\alpha\alpha}^{-}$; black dashed diamonds denote $P_{\alpha\alpha}$. The vertical dashed line marks charge neutrality. (c) Polar plot of the longitudinal conductivity polarization $P(\theta)$ as a function of the in-plane current angle $\theta$ over $0^\circ$--$360^\circ$.}
\label{figure4}
\end{figure*}

Once SOC locks the N\'eel axis to the lattice and the relevant unitary mirror is removed, the leading local Hamiltonian admits both displacement and mass terms,
\begin{equation}
\begin{aligned}
H_i(\mathbf q)=&\,(\varepsilon_{W,i}^{(0)}+\delta_i)\sigma_0
+[v_{i,\parallel}q_{\parallel}+a_{i,\parallel}]\sigma_x\\
&+[v_{i,\perp}q_{\perp}+a_{i,\perp}]\sigma_z
+m_i\sigma_y .
\end{aligned}
\label{eq:general_kp}
\end{equation}
Here $a_{i,\parallel}$ and $a_{i,\perp}$ shift the cone in momentum, $\delta_i$ shifts its energy, and $m_i$ controls the local gap. Minimizing first along the parent mirror-invariant line and then over the full two-dimensional momentum plane gives
\begin{equation}
\begin{gathered}
\Delta_{\rm path}^{(i)}=2\sqrt{a_{i,\perp}^{\,2}+m_i^2},
\qquad
\Delta_{\rm 2D}^{(i)}=2|m_i|,\\
q_{\parallel,i}^{*}=-a_{i,\parallel}/v_{i,\parallel},
\qquad
q_{\perp,i}^{*}=-a_{i,\perp}/v_{i,\perp}.
\end{gathered}
\label{eq:path_vs_true_gap}
\end{equation}
Equation~(\ref{eq:path_vs_true_gap}) separates the path-restricted splitting from the local two-dimensional gap \cite{PhysRevLett.121.106402}. A transverse displacement can produce a finite $\Delta_{\rm path}$ even when $\Delta_{\rm 2D}$ vanishes. A detailed derivation of the path-restricted and two-dimensional gaps, including the displaced-node limit and node-energy inequivalence, is given in the SM~\cite{SM}.

For an in-plane N\'eel vector, the combined symmetry
\(\Theta=C_{2z}\mathcal T\) maps every in-plane momentum back onto itself and satisfies \(\Theta^2=+1\). This symmetry forbids the coupling needed to open a gap at an isolated Weyl crossing, but it does not pin the crossing to its original momentum. The Weyl point can therefore move away from the parent high-symmetry line while remaining gapless. Consequently, the band structure along the original high-symmetry path may appear gapped even though the crossing survives nearby. The Weyl point disappears only if it annihilates with another one or if \(C_{2z}\mathcal T\) is broken. Our full-Brillouin-zone calculations confirm that the present reconstruction is dominated by such a momentum shift.

{What remains to explain is why the nodes protected by the surviving antiunitary symmetry are no longer pinned to the same energy. In the parent $C_{4v}$ phase, the two mirror lines $X\!-\!M_1$ and $X\!-\!M_3$ are related by the unitary twofold rotation $C_{2z}$. This symmetry enforces $\delta_1=\delta_3$ and hence the equivalence of the two cones. Breaking unitary $C_{2z}$ lifts the constraint $\delta_1=\delta_3$ and allows independent scalar shifts $\delta_{1,3}$, i.e., $E_W^{(1)}-E_W^{(3)}=\delta_1-\delta_3\neq0$, so the two cones are no longer equivalent. The antiunitary $C_{2z}\mathcal T$ forbids an independent local mass, whereas breaking the unitary $C_{2z}$ symmetry lifts the equivalence between the two surviving Weyl cones, providing the second ingredient of the spin-filtering mechanism.}

The high-symmetry path and corresponding band structures show which crossings remain pinned by the surviving spatial mirrors, whereas full-zone analysis resolves momentum displacement and local gap. Thus, $\mathbf{n}\parallel\hat{x}$ preserves the spatial mirror $M_x$ but breaks $M_y$. Hence, once the corresponding spatial mirror is lost, the orthogonal parent pair is no longer positionally constrained. One mirror remains a spatial symmetry and protects one set of crossings, while the other becomes magnetic. Selective mirror breaking explains why only the protected set remains visible on the conventional path. When $\mathbf n$ lies along $x$ or $y$, one in-plane mirror remains a spatial symmetry while the orthogonal one becomes magnetic, and tetragonal symmetry exchanges the in-plane axes and associated spin projections. For $\mathbf n$ along the diagonal $d\equiv(110)$, neither in-plane mirror remains a spatial symmetry, and the spatial rotation $C_{2z}$ is also absent, so path-resolved branches may be separated and energy inequivalent. The representative configurations illustrate how the N\'eel-vector orientation governs symmetry and, consequently, the nature of the path-resolved branches. However, the spectra do not distinguish displacement from a true mass gap.

The full-BZ Wannier analysis makes this distinction [Figure~\ref{figure3}(d)]. It identifies two boundary nodes, $W_1$ and $W_2$, with loop Berry phases close to $\pi$. It also resolves two off-symmetry Weyl-derived cones, $O_1$ and $O_2$, displaced from the parent high-symmetry line and lying within a few meV of charge neutrality. Details of the Wannier construction and interpolation benchmark, together with the full-BZ node search, refined coordinates, residual splittings, and Berry-phase calculations, are provided in the SM~\cite{SM}.

The mirror-pinned $W_1$ and $W_2$ nodes cease to be isoenergetic because $\mathbf n\parallel x$ breaks the unitary twofold rotation $C_{2z}$. In the calculated spectrum, these nodes lie on opposite sides of the charge-neutral chemical potential, and their electron- and hole-like branches form finite Fermi-surface pockets in one spin-projected manifold\cite{PhysRevB.85.165110}. In the opposite manifold, loss of the unitary-mirror constraint displaces the $O_1$ and $O_2$ Weyl-derived cones from their parent high-symmetry lines to generic momenta, leaving substantially less low-energy phase space near charge neutrality. A metallic spin-projected manifold therefore coexists with a low-DOS Weyl-derived manifold, constituting the metal--Weyl-semimetal state, as shown in Figure \ref{figure3}(e).

To determine whether this spectral imbalance survives velocity weighting, we calculated the longitudinal sheet conductance in the constant-relaxation-time Boltzmann approximation using the same spinor Wannier Hamiltonian \cite{MADSEN200667,PIZZI2014422}.
Because SOC mixes spin, \(G_{\alpha\alpha}^{\pm}\) denotes the contribution projected onto the two eigenvalues of the spin operator along the magnetic axis, rather than two independently conserved transport channels.
We evaluate \(P_{\alpha\alpha}\) using the first relation in Eq.~(\ref{eq:transport-decomposition}), while the second relation separates the spectral-weight and velocity-moment factors controlling the conductance imbalance.
The resulting projected conductances and longitudinal polarizations are shown in Figs.~\ref{figure4}(a) and \ref{figure4}(b).
The absolute conductances scale with the common relaxation time, whereas \(P_{\alpha\alpha}\) is independent of its chosen value under the equal-\(\tau\) assumption.

At charge neutrality and $20$~K, direct angular full-BZ integrations after projecting the band velocity onto each sampled current direction give positive longitudinal polarization for $\mathbf n\parallel x$ at every one of the 72 sampled in-plane current directions [Figure~\ref{figure4}(c)]. The polarization ranges from $P_{xx}=76.4\%$ for current along $x$ to $P_{yy}=82.0\%$ along $y$. The angular profile is mirror symmetric about the $x$ axis, consistent with the surviving unitary mirror $M_x$. When $\mathbf n$ lies along $x$ or $y$, tetragonal symmetry exchanges the in-plane axes and the associated altermagnetic spin projections.

In summary, we introduce a N\'eel-vector-oriented framework for magnetic space-group reduction that enables symmetry-selective Weyl-cone reconstruction in two-dimensional altermagnets. Using monolayer Ta$_2$TeSeO as a prototype, we show that without SOC, each spin subspace hosts a pair of mirror-protected Weyl points. Fixing the N\'eel vector along \(x\) (\(y\)) preserves the unitary mirror \(M_x\) (\(M_y\)) while converting its orthogonal partner into a magnetic mirror, allowing the affected Weyl cones to move away from their original high-symmetry line. Moreover, breaking the unitary \(C_{2z}\) symmetry lifts the energy equivalence between the two surviving Weyl cones, while the retained \(C_{2z}\mathcal T\) symmetry prevents a gap from opening at the crossings but allows their positions to shift. Full-Brillouin-zone Wannier searches and direct VASP calculations confirm that these shifted cones remain gapless. Near charge neutrality, low-DOS Weyl states therefore coexist with metallic states having a much larger DOS. For \(\mathbf n\parallel x\) at 20~K, the longitudinal conductivity polarization ranges from \(76.4\%\) to \(82.0\%\) for all in-plane current directions. Because the two spin projections have comparable band velocities, this transport imbalance arises mainly from their different DOS values rather than from strongly anisotropic velocity filtering. 

\begin{acknowledgments}
We thank Xiangru Kong for helpful discussions. We would like to thank Shandong Institute of Advanced Technology and National Supercomputing Center (Shuguang) for providing computational resources. This work is supported by the National Natural Science Foundation of Shandong Province (Grant No.~ZR2024QA040). Xin Chen thanks the China Scholarship Council for financial support (No.~201606220031). Duo Wang acknowledges financial support from the Science and Technology Development Fund of Macao SAR (Nos.~0062/2023/ITP2 and 0016/2025/RIA1) and Macao Polytechnic University (Grant No.~RP/FCA-03/2023). Biplab Sanyal acknowledges financial support from the Swedish Research Council (Grant No.~2022-04309) and the STINT Mobility Grant for Internationalisation (Grant No.~MG2022-9386).
\end{acknowledgments}

\bibliography{Ref}

\end{document}